\documentclass[aoas,nameyear,seceqn,dvips]{arximspdf}
\usepackage{dcolumn}
\usepackage{graphicx}

\doi{10.1214/09-AOAS317}
\volume{4}
\issue{3}
\pubyear{2010}
\firstpage{1291}
\lastpage{1310}

\makeatletter
\newcolumntype{d}[1]{D{.}{.}{#1}}
\makeatother

\begin{document}
\begin{frontmatter}

\title{Assessing the protection provided by misclassification-based
disclosure limitation methods for survey microdata}
\runtitle{Assessing the protection provided by misclassification}

\begin{aug}
\author[a]{\fnms{Natalie} \snm{Shlomo}\ead[label=e2]{N.Shlomo@Soton.ac.uk}} and
\author[a]{\fnms{Chris} \snm{Skinner}\corref{}\ead[label=e1]{C.J.Skinner@Soton.ac.uk}}
\runauthor{N. Shlomo and C. Skinner}
\affiliation{University of Southampton}
\address[a]{School of Social Sciences\\
University of Southampton\\
Southampton, SO17 1BJ \\United Kingdom\\
\printead{e1}\\
\phantom{E-mail: }\printead*{e2}} 
\end{aug}

\received{\smonth{10} \syear{2008}}
\revised{\smonth{11} \syear{2009}}

%
\begin{abstract}
Government statistical agencies often apply statistical disclosure
limitation techniques to survey microdata to protect the
confidentiality of respondents. There is a need for valid and practical
ways to assess the protection provided. This paper develops some simple
methods for disclosure limitation techniques which perturb the values
of categorical identifying variables. The methods are applied in
numerical experiments based upon census data from the United Kingdom
which are subject to two perturbation techniques: data swapping (random
and targeted) and the post randomization method. Some simplifying
approximations to the measure of risk are found to work well in
capturing the impacts of these techniques. These approximations provide
simple extensions of existing risk assessment methods based upon
Poisson log-linear models. A numerical experiment is also undertaken to
assess the impact of multivariate misclassification with an increasing
number of identifying variables. It is found that the misclassification
dominates the usual monotone increasing relationship between this
number and risk so that the risk eventually declines, implying less
sensitivity of risk to choice of identifying variables. The methods
developed in this paper may also be used to obtain more realistic
assessments of risk which take account of the kinds of measurement and
other nonsampling errors commonly arising in surveys.
\end{abstract}

%

\end{frontmatter}

\section{Introduction}
Government statistical agencies have statutory and ethical obligations
to protect the confidentiality of the data they collect. At the same
time, their core mission is to ensure that these data are used
effectively for statistical purposes. Tensions between these two
objectives may arise, in particular, when access to microdata on
individuals or establishments is to be provided to researchers, so that
they may conduct their own analyses of social or economic phenomena.
Although microdata may be anonymized by removing obvious identifying
information such as name and address without damage to the statistical
analyses, such anonymization will rarely be considered sufficient for
confidentiality protection, since the rich socio-economic information
in the microdata may often enable records to be identified by matching
to another data source on known individuals or establishments. Agencies
have therefore developed a number of ways of protecting confidentiality
in this context. One common approach is to modify the microdata file by
applying a statistical disclosure limitation (SDL) method, such as
recoding or data perturbation, to those variables judged potentially
identifying [Federal Committee on Statistical Methodology (\citeyear{FCSM2005})]. Such
modification can, however, seriously reduce the utility of the
microdata and it is therefore important for the agency to be able to
assess the protection provided by such methods in order to be able to
make judgements about the degree of modification to apply.

The aim of this paper is to develop methodology to assess the
disclosure protection provided by the misclassification of one or more
categorical identifying variables. Misclassification is supposed here
to arise in one of two ways. First, it may be the result of the
deliberate application by the agency of an SDL method, specifically we
consider the methods of data swapping [Dalenius and Reiss (\citeyear{DaleniusReiss1982})] and
post-randomization or PRAM [Gouweleeuw et al. (\citeyear{GouweleeuwEtAl1998})]. This paper is
motivated by experience of the use of such methods at government
statistical agencies (especially in the United Kingdom) with microdata
from social surveys on individuals or from population censuses. In
these cases, the potential identifying variables which might be used
for matching are invariably categorical. A second way in which
misclassification may arise is as a result of measurement error which
arises naturally in surveys and takes the form of misclassification for
categorical variables [Kuha and Skinner (\citeyear{KuhaSkinner1997})]. In this case, we shall
suppose that the agency has some information about the nature of the
misclassification mechanism.

In the current practice of statistical agencies, when the disclosure
protection of such methods is assessed, it is usually based upon simple
measures, such as functions of the diagonal elements of the
misclassification matrix [Willenborg and De Waal (\citeyear{WillenborgDeWall2001}), page~119], or
a simple estimated probability that an apparent match is correct
[Gouweleeuw et al. (\citeyear{GouweleeuwEtAl1998})], or via the outcome of a record linkage
experiment (see below). Reiter (\citeyear{Reiter2005}) developed a more sophisticated
approach by defining a measure of identification risk, based upon the
modeling framework of Duncan and Lambert (\citeyear{DuncanLambert1989}), and showing how it
could be assessed before and after the application of a number of SDL
methods, including data swapping. This focus on identification risk is
often appropriate in government contexts, where judgments about
protection are informed by legislation or codes of practice which
express threats to confidentiality in terms of individual respondents
being identified. However, the need to model a very wide range of
microdata variables and relationships in Reiter's (\citeyear{Reiter2005}) approach may
limit its application in practice. In this paper we develop an approach
which is based on a similar framework to Reiter (\citeyear{Reiter2005}), but which
retains some of the simplicity of the former methods. We achieve
simplification by restricting the information set upon which the risk
measure is conditioned, extending the approach of Skinner and Shlomo (\citeyear{SkinnerShlomo2008}).
Our approach also extends Reiter (\citeyear{Reiter2005}) by taking fuller
account of the protection achieved from sampling.

Assessing identification risk using record linkage experiments [e.g.,
Yancy, Winkler and Creecy (\citeyear{YanceyWinklerCreecy2002});
Domingo-Ferrer and Torra (\citeyear{DomingoferrerTorra2001})] is
natural given the threat that such methods pose [Fienberg (\citeyear{Fienberg2006})]. The
experiment typically involves matching records in the microdata file,
masked by an SDL method, to records in the original unmasked file. The
risk is often defined as the proportion of such matches which are
correct [Spruill (\citeyear{Spruill1982})]. A problem with this approach is that it makes
an unjustified assumption that a hypothetical intruder has access to
data that are as good as the original data and may not take account of
the disclosure protection provided by sampling. We shall show in the
\hyperref[appendix]{Appendix} that our proposed approach to assessing identification risk in
the case of exact matching does, in fact, provide a closed form
expression for the correct match proportion which would be estimated by
an experiment using a form of probabilistic record linkage proposed by
Fellegi and Sunter (\citeyear{FellegiSunter1969}). Record linkage experiments have the
potential to capture the impact of a wider range of types of potential
attack, including those that make explicit allowance for data masking
and exploit greater computational power [Winkler (\citeyear{Winkler2004})], but
consideration of such extensions is beyond the scope of this paper.

Statistical modeling approaches to identification risk assessment have
been proposed by a number of authors [e.g., Paass (\citeyear{Paass1988});
Duncan and Lambert (\citeyear{DuncanLambert1989}); Fuller (\citeyear{Fuller1993})]. It is generally assumed that an
intruder seeks to identify an individual in the microdata by matching
records to known individuals in the population using identifying
variables, also called key variables, values of which are known both
for the microdata records and for the known individuals. This paper
builds on the literature which has used models for categorical key
variables as a basis for assessing disclosure risk.
Bethlehem, Keller and Pannekoek (\citeyear{BethlehemKellerPannekoek1990}) is a seminal contribution. We follow especially
Skinner and Shlomo (\citeyear{SkinnerShlomo2008}), who considered the use of log-linear models to assess
identification risk. Their work did not, however, consider the impact
of SDL methods on risk, other than the recoding of key variables.

The empirical work in this paper is based upon the 2001 population
census in Great Britain, which will be used to provide population data
to validate risk assessments for samples, viewed as representing
potential sample surveys. Our focus will be on the impact of SDL
methods on identification risk. The effects of these methods on the
utility of potential data analyses is also vitally important and we
provide some information loss measures to analyze and compare the
perturbation methods.

Our paper is organized as follows. Measures of identification risk in
the presence of misclassification are developed in Section~\ref{za}.
Since these measures depend upon population quantities which may be
unknown, methods of estimating these measures using sample data are
considered in Section~\ref{zb}. Applications using census data are
presented in Section~\ref{zc} for a random and targeted data swapping
method and a random and targeted post-randomization method (PRAM). A
further numerical illustration with multivariate misclassification is
presented in Section~\ref{zq}. Section~\ref{zs} contains a concluding
discussion.

\section{Identification risk under misclassification}\label{za}

Consider the release of a microdata file consisting of records for a
sample $s=\{1,2,\ldots,n\}$ drawn from a finite population $U$ of size
$N$. We suppose an intruder seeks to match a known target unit in $U$
to a record in the file using $C$ categorical key variables $X^1,\ldots
,X^C$. We assume the agency knows the intruder's choice of key
variables. Possible departures from this assumption are discussed in
Section~\ref{zs}. The variable formed by cross-classifying the key variables,
as measured by the intruder on the target unit, is denoted $X$ and its
values are labeled $1,2,\ldots,K$. The value of $X$ recorded in the
microdata, after the application of the SDL method (and natural
measurement error), is denoted $\tilde{X}$. We treat the values of $X$
for population units as fixed and suppose the values of $\tilde{X}$ for
the records in the microdata are determined independently by a
misclassification matrix $M$, where
%
\begin{equation}
\operatorname{Pr}(\tilde{X}=j|X=k)=M_{jk}. \label{ab}
\end{equation}

To assess the disclosure protection provided by misclassification, we
imagine that the intruder observes a match between a specific sample
unit $A$ and a target population unit $B$, that is, observes $\tilde
{X}_{A}=X_{B}$ (where $\tilde{X}_{A}$ is the value of $\tilde{X}$ for
unit $A$ and $X_{B}$ is the value of $X$ for unit $B$), and measures
disclosure risk in terms of the uncertainty as to whether $A=B$. A
simple ad hoc measure of this uncertainty is given by $M_{jj}$ (or
$1-M_{jj}$), where $j$ is the common value of $\tilde{X}_{A}$ and
$X_{B}$. Willenborg and De Waal [(\citeyear{WillenborgDeWall2001}), page~121] propose that the
agency specifies upper bounds for these diagonal elements of $M$
according to the level of protection required. Following Reiter (\citeyear{Reiter2005}),
we define the identification risk as $\operatorname{Pr}(A=B|\mathit{data})$, where the values
$\tilde{X}_{A}$ and $X_{B}$ are implicitly included in the $\mathit{data}$ and
the nature of the probability mechanism will be clarified later. A
simplified approach to estimating this risk is given by
Gouweleeuw et al. (\citeyear{GouweleeuwEtAl1998}), who make the very
conservative assumption that the intruder knows that $B$ is in the
sample and approximate $\operatorname{Pr}(A=B|\mathit{data})$ by $\operatorname{Pr}(X_{A}=j|\tilde
{X}_{A}=j)=M_{jj}\operatorname{Pr}(X_{A}=j)/\sum_{k}M_{jk}\operatorname{Pr}(X_{A}=k)$, which they
estimate by
%
\begin{equation}
M_{jj}f_{j}\Big/\sum_{k}M_{jk}f_{k}, \label{ht}
\end{equation}
where $f_{k}$ is the number of units in $s$ for which $X=k$ (they in
fact use the odds rather than the probability). In contrast to the
highly simplifying assumptions of Gouweleeuw et al. (\citeyear{GouweleeuwEtAl1998}),
Reiter (\citeyear{Reiter2005}) allows for considerable generality by adopting a very wide
definition of $\mathit{data}$ in $\operatorname{Pr}(A=B|\mathit{data})$, so that it may include all the
values of $\tilde{X}_{i}$ in the sample as well as the values of any
other microdata variables. This creates not only a major modeling task
to assess the probability of interest, but also the possibility that
this probability will be sensitive to the specification of the model.

We seek an intermediate position, avoiding the very conservative
assumption that the intruder knows that $B$ is in the sample, but
reducing the scope of $\mathit{data}$ in $\operatorname{Pr}(A=B|\mathit{data})$ to avoid the complex
modeling issues. We define the matching variable $\tilde{Z}_{i}$ to be
1 if $\tilde{X}_{i}=X_{B}$ and 0 otherwise and we take the data to
consist of the values $\tilde{Z}_{i}$ for $i\in{s}$. We suggest that
this is the critical information to consider when assessing the
probability that an observed match is correct. We shall also restrict
our attention further to the case when a unique sample unit matches $B$
(so $\tilde{Z}_{a}=1$ and $\tilde{Z}_{i}=0$ if $i\neq{a}$ for some unit
$a\in{s}$). This is the worst case and thus of most interest, that is,
the risk will be lower if $B$ matches more than one sample unit. In
this case, we obtain the following expression for the identification risk:
\begin{eqnarray}\label{cd}
\mathit{Identification\ risk}&=&\operatorname{Pr}(A=B|\tilde{Z}_{1},\ldots,\tilde{Z}_{n})\nonumber \\[-8pt]\\[-8pt]
&=& \operatorname{Pr}(E_{B})\Big/\sum_{a\in{U}}\operatorname{Pr}(E_{a}),\nonumber
\end{eqnarray}
where $E_{a}$ is the event that population unit $a$ is sampled and its
value $\tilde{X}_{a}$ matches $X_{B}$ and that no other population unit
is both sampled and has a value of $\tilde{X}$ which matches $X_{B}$.
In order to allow for the effect of unequal probability sampling and
the potential use of sampling weights, we suppose that units in the
population $U$ are selected independently into the sample $s$ with
inclusion probabilities $\pi_{j}$ which may depend on the value $\tilde
{X}=j$ for the unit. Writing $X_{a}=k$ and $X_{B}=j$ and using our
previous assumptions about the misclassification mechanism, we obtain
$\operatorname{Pr}(E_{a})=\alpha_{j}{M_{jk}}/(1-\pi_{j}{M_{jk}})$, where $\alpha
_{j}=\pi_{j}\prod_{l}(1-\pi_{j}{M_{jl}})^{F_{l}}$ and $F_{j}$ is the
number of units in the population with $X=j$. Hence,%
\begin{eqnarray}\label{ef}
&&\operatorname{Pr}(A=B|\tilde{Z}_{1},\ldots,\tilde{Z}_{n})\nonumber
\\[-8pt]\\[-8pt]
&&\qquad=[M_{jj}/(1-\pi_{j}{M_{jj}})]
\Big/\biggl[\sum_{k}F_{k}M_{jk}/(1-\pi_{j}{M_{jk}})\biggr].\nonumber
\end{eqnarray}
This expression assumes the intruder does not know whether $B\in{s}$.
If this event was known to arise and was included in the conditioning
set, (\ref{ef}) should be modified by setting $\pi_{j}=1$ and replacing
$F_{k}$ by $f_{k}$. This produces an expression that is similar to that
given earlier in (\ref{ht}) from Gouweleeuw et al. (\citeyear{GouweleeuwEtAl1998}) but makes
fewer approximations. For expression (\ref{ef}), the identification
risk also assumes that the $F_{k}$ are part of the data, that is,
known. In practice, this will often not be the case, as discussed by
Skinner and Shlomo (\citeyear{SkinnerShlomo2008}), and it will be necessary to integrate the
$F_{k}$ out of this expression as will be discussed in Section~\ref{zb}. It
follows from (\ref{ef}) that
\[
\operatorname{Pr}(A=B|\tilde{Z}_{1},\ldots,\tilde{Z}_{n})\leq{1/F_{j}}
\]
with equality holding if there is no misclassification. The extent to
which the left-hand side of this inequality is less than the right-hand
side measures the impact of misclassification on disclosure risk.

If the inclusion probabilities $\pi_{j}$ are all small, we may
approximate (\ref{ef}) by
\[
\operatorname{Pr}(A=B|\tilde{Z}_{1},\ldots,\tilde{Z}_{n})= M_{jj}\Big/\biggl(\sum_{k}F_{k}M_{jk}\biggr).
\]

Moreover, if the population size is large, we have approximately
$\sum_{k}F_{k}M_{jk}\approx\tilde{F}_{j}$, where $\tilde
{F}_{j}$ is the number of units in the population which would have
$\tilde{X}=j$ if they were included in the microdata (with
misclassification). Hence, a simple approximate expression for the
risk, natural for many social surveys, is%
\begin{equation}\label{gh}
\operatorname{Pr}(A=B|\tilde{Z}_{1},\ldots,\tilde{Z}_{n})= M_{jj}/\tilde{F}_{j}.
\end{equation}

An alternative derivation of this result is provided in the \hyperref[appendix]{Appendix}
under the assumption that the intruder adopts the probabilistic record
linkage approach of Fellegi and Sunter (\citeyear{FellegiSunter1969}), making a link if the
match variable $\tilde{Z}_{a}=1$. The identification risk is the
probability that the match is correct and the above approximation is
obtained if the probability is defined with respect to the sampling
scheme, the misclassification mechanism and a random selection of a
pair for matching as in Fellegi and Sunter (\citeyear{FellegiSunter1969}).

Another approximation to expression (\ref{ef}) is obtained by assuming
the misclassification is small, say, $M_{jj}=(1-\delta)\phi_{jj}$ and
$M_{jk}=\delta\phi_{jk}$ $(j\neq{k})$, where the $\phi$ are fixed and
$\delta\stackrel{}\rightarrow0$. In this case, we have%
\begin{eqnarray}\label{ij}
&&\operatorname{Pr}(A=B|\tilde{Z}_{1},\ldots,\tilde{Z}_{n})\nonumber
\\[-8pt]\\[-8pt]
&&\qquad\approx{F}_{jj}^{-1}\bigl(1-[\tilde{F}_{j}-F_{j}M_{jj}]/[F_{j}M_{jj}/(1-\pi_{j}{M}_{jj})]\bigr)\nonumber
\end{eqnarray}
or
\begin{eqnarray}\label{kl}
&&\operatorname{Pr}(A=B|\tilde{Z}_{1},\ldots,\tilde{Z}_{n})\nonumber
\\[-8pt]\\[-8pt]
&&\qquad\approx[M_{jj}/(1-\pi_{j}{M}_{jj})]/[(F_{j}\pi_{j}{M}_{jj}^{2})/(1-\pi{M}_{jj})+\tilde{F}_{j}].\nonumber
\end{eqnarray}

Note that none of approximations (\ref{gh}), (\ref{ij}) or (\ref{kl})
depend upon $M_{jk}$ for $j\neq{k}$ and so knowledge of these
probabilities is not required in the estimation of risk.

The definition of risk in (\ref{cd}) applies to a specific record.
Agencies will also usually wish to consider aggregate measures to
enable them to make judgements about the whole file. Following
Skinner and Shlomo (\citeyear{SkinnerShlomo2008}), we define an aggregate measure as the sum of the
record-level measures in (\ref{ef}) across sample unique records:
%
\begin{equation}
\tau=\sum_{j\in{SU}}[M_{jj}/(1-\pi_{j}{M_{jj}})]\Big/\biggl[\sum
_{k}F_{k}M_{jk}/(1-\pi_{j}{M_{jk}})\biggr],
\label{vt}
\end{equation}
where $SU$ is the set of key variable values which are sample unique.
This measure may be interpreted as the expected number of correct
matches among sample uniques.
For some purposes, an agency might find this measure easier to
interpret if it is transformed into a measure with an upper bound, such
as by dividing by the number of sample uniques to obtain a proportion.
However, we shall stick with the untransformed $\tau$ as a measure of
the total number of units, for example, individuals, threatened with
identification.

We also consider, for comparison, a related measure which could be used
if the misclassification status of microdata records is known. Let
$\mathit{SUCC}$ denote the set of key variable values which are sample unique
and where these sample unique values have been correctly classified.
The measure is given by%
\begin{equation}
\tau^{*}_{\mathit{CC}}=\sum_{j\in{\mathit{SUCC}}}1/F_{j},\label{uv}
\end{equation}
and again may be interpreted as the expected number of correct matches
among sample uniques. We also define $\tau^{*}$ as the corresponding
measure of risk in the absence of perturbation, that is, the sum of
$1/F_{j}$ across key values which are unique in the sample with respect
to $X$.

\section{Risk estimation}\label{zb}

An agency wishing to apply an SDC method to survey microdata will
generally not know the values of $F_{j}$ or $\tilde{F}_{j}$ appearing
in the risk expressions. We do suppose that the values of $M_{jk}$ are
known. Skinner and Shlomo (\citeyear{SkinnerShlomo2008}) discuss the estimation of risk in the
absence of misclassification based on a Poisson log-linear model. In
this case, expression (\ref{ef}) reduces to $1/F_{j}$ and their broad
approach is to define the risk as the conditional expectation of this
quantity given the observed data and to estimate this expectation using
data for the sample counts $f_{j}$, $j=1,2,\ldots,K$, for which a
log-linear model is fitted. Expression (\ref{gh}) provides a simple way
to extend their approach to misclassification provided $M_{jj}$ is
known. Since the $\tilde{f}_{j}$, $j=1,2,\ldots,K$, represent the
available data, all that is required is to ignore the misclassification
and estimate the expectation of $1/\tilde{F}_{j}$ given the data from
the $\tilde{f}_{j}$, $j=1,2,\ldots,K$, as in Skinner and Shlomo (\citeyear{SkinnerShlomo2008}),
that is, by fitting a log-linear model now to the $\tilde{f}_{j}$,
$j=1,2,\ldots,K$, following the same criteria as before. This results
in an estimate $\hat{E}(1/\tilde{F}_{j}|\tilde{f}_{j}=1)$ based on the
assumptions of the Poisson distribution for the population and sample
counts. These estimates should be multiplied by the $M_{jj}$ values and
summed if aggregate measures of the form in (\ref{vt}) are needed. It
would appear to be rather more complex to estimate the expressions
including terms in $F_{j}$. In the presence of complex sampling, the
estimation method may be adapted using the method of pseudo maximum
likelihood estimation [Rao and Thomas (\citeyear{RaoThomas2003})] by incorporating survey
weights in the estimation as discussed by Skinner and Shlomo (\citeyear{SkinnerShlomo2008}).

\section{Application of perturbative disclosure limitation techniques}\label{zc}

In this section we consider two specific perturbative SDL techniques
used at statistical agencies: data swapping and the post-randomization
method (PRAM). Both techniques introduce misclassification of the key
variables to lower the probabilities of identifying individuals. We
present examples of how to assess the impact of these techniques on
identification risk. Since the misclassification is under the control
of the statistical agency, the misclassification matrix $M$ is known.
\subsection{Data swapping}\label{ss}

The method of data swapping is based on exchanging the values of one or
more key variables between pairs of records. In order to minimize bias,
the pairs of records are typically selected within strata defined by
control variables, such as the age and sex of the individual. In
addition, the perturbation can be targeted to high-risk records that
are more likely to be population uniques, for example, on rare
ethnicities. It is common that geographic variables are swapped between
records for the following reasons:
\begin{itemize}
\item For given values of the control variables, the sensitive
variables are likely to be relatively independent of geography and,
therefore, it is expected that less bias will occur. In addition,
swapping geography will not normally result in inconsistent and
illogical records. By contrast, swapping a variable such as age would
result in many inconsistencies with other variables, such as marital
status and education.
\item At a higher geographical level and within control strata, the
marginal distributions are preserved.
\item The level of protection increases by swapping variables which
are highly ``matchable'' such as geography.
\end{itemize}

For this experiment, we carry out a simple data swapping procedure
where the geography variable of Local Authority District (LAD) is
exchanged between a pair of individuals. The population includes
$N=1{,}468{,}255$ individuals from an extract of the 2001 United Kingdom
(UK) Census. We drew $1\%$ Bernoulli samples ($n=14{,}683$) and define
six key variables for the risk assessment: Local Authority (LAD) (11),
sex (2), age groups (24), marital status (6), ethnicity (17), economic
activity (10), where the numbers of categories of each variable are in
parentheses ($K=538{,}560$). We implement a random data swap by drawing a
sub-sample of $10\%$ and $20\%$ in each of the LADs. The remaining
individuals are not perturbed. On the sub-samples in each LAD, half of
the individuals are flagged. For each flagged individual, an unflagged
individual is randomly chosen within the sub-sample and their LAD
variables swapped, on condition that the individual chosen was not
previously selected for swapping and that the two individuals do not
have the same LAD, that is, no individual is selected twice for
producing a pair. We also implemented a $10\%$ and $20\%$ targeted data
swap where the LAD variable is swapped separately within two groups
defined by ``White British'' and ``Other'' ethnicities. For the $20\%$
swap, LADs were swapped randomly between all pairs of individuals in
the ``Other'' group and a small percentage ($7\%$) of individuals in the
``White British'' group. This swapping rate was chosen so that the total
percentage of swapped individuals would be $20\%$ as in the random data
swapping. For the $10\%$ swap, LADs were swapped randomly from among
the ``Other'' group that compose $10\%$ of the total individuals in the sample.

The misclassification matrix $M$ for the data swapping designs can be
expressed simply in terms of the $11\times11$ misclassification matrix,
denoted $M^{g}=[M^{g}_{jk}]$, for the LAD variable $g$:

For the random swap:
\begin{itemize}
\item On the diagonal: $M^{g}_{jj}=0.9$ or $M^{g}_{jj}=0.8$ for the
$10\%$ and $20\%$ swaps respectively.
\item Off the diagonal:\vspace*{1pt} $M^{g}_{jk}=0.1\times{n_{k}}/(\sum_{l\neq
{j}}n_{l})$ or $M^{g}_{jk}=0.2\times{n_{k}}/(\sum_{l\neq{j}}n_{l})$,
where $n_{k}$ is the number of records in the sample in LAD $k$,
$k=1,2,\ldots,11$, for the $10\%$ and $20\%$ swaps respectively.
\end{itemize}

For the targeted swap on the $10\%$ swap, the values $M^{g}_{jk}$ for
the ``Other'' ethnicity are calculated as follows:
\begin{itemize}
\item On the diagonal: $M^{g}_{jj}=0.25$.
\item Off the diagonal: $M^{g}_{jk}=0.75\times{n_{2k}}/(\sum_{l\neq
{j}}n_{2l})$, where $n_{2k}$ is the number of records in the sample
with ``Other'' ethnicity in LAD $k$, $k=1,2,\ldots,11$.
\end{itemize}

For the targeted swap on the $20\%$ swap, the misclassification matrix
$M$ is defined separately according to the ``White British'' and ``Other''
ethnicities as follows:
\begin{itemize}
\item On the diagonal: $M^{g}_{jj}=0.93$.
\item Off the diagonal: $M^{g}_{jk}=0.07\times{n_{1k}}/(\sum_{l\neq
{j}}n_{1l})$, where $n_{1k}$ is the number of records in the sample
with ``White British'' ethnicity in LAD $k$, $k=1,2,\ldots,11$.
\end{itemize}
The values $M^{g}_{jk}$ for the ``Other'' ethnicity are calculated as follows:
\begin{itemize}
\item On the diagonal:\vspace*{2pt} $M^{g}_{jj}=0$.
\item Off the diagonal:\vspace*{2pt} $M^{g}_{jk}=1\times{n_{2k}}/(\sum_{l\neq
{j}}n_{2l})$, where $n_{2k}$ is the number of records in the sample
with ``Other'' ethnicity in LAD $k$, $k=1,2,\ldots,11$.
\end{itemize}

\subsection{The post-randomization method (PRAM)}

A more direct method that is used for exchanging values of categorical
variables is PRAM. For this method, values of categories in a given
record are changed or not changed stochastically according to a
misclassification matrix. This matrix is chosen to preserve expected
marginal frequencies of the variables. Let $f^{c}$ be the row vector of
sample frequencies of the different categories of key variable $X^{c}$
and $p^{c}=f^{c}/n$ be the corresponding vector of sample proportions,
where $n$ is the sample size. For each record, the category of $X^{c}$
is changed or not changed according to the probabilities in the
misclassification matrix $M^{c}$. Let $\tilde{f}^{c}$ be the row vector
of perturbed frequencies. Then $E(\tilde{f}^{c}|f^{c})=f^{c}M^{c}$,
where the expectation is with respect to the misclassification
mechanism. The matrix $M^{c}$ may be expected to be nonsingular since
small perturbation rates should imply that it is ``close to'' diagonal.
The inverse $M^{{c}^{-1}}$ can be used to obtain an unbiased estimator
of the original frequency vector: $\hat{f}^{c}=\tilde
{f}^{c}M^{{c}^{-1}}$. In addition, we can place the condition of
invariance on the matrix $M^{c}$, that is, $f^{c}M^{c}=f^{c}$, and
preserve the expected marginal frequencies. This releases the users of
the perturbed file of the extra effort to obtain unbiased moment
estimates of the original data, since $\tilde{f}^{c}$ itself will be an
unbiased estimate of $f^{c}$.

To obtain an invariant transition matrix, the following two-stage
algorithm is applied [see Willenborg and De Waal (\citeyear{WillenborgDeWall2001})]. Let $M^{c}$
be the misclassification matrix: $M^{c}_{jk}=\operatorname{Pr}(\tilde
{X}^{c}=k|X^{c}=j),$ where $j$ represents the original category and $k$
the perturbed category. Now calculate the matrix $Q$ using the Bayes
formula by $Q^{c}_{kj}=\operatorname{Pr}(X^{c}=j|\tilde
{X}^{c}=k)=M^{c}_{jk}\operatorname{Pr}(X^{c}=j)/[\sum_{l}M^{c}_{lk}\operatorname{Pr}(X^{c}=l)]$. We
estimate the entries of this matrix by $\hat
{Q}^{c}_{kj}=M^{c}_{jk}p^{c}_{j}/[\sum_{l}M^{c}_{lk}p^{c}_{l}]$, where
$p^{c}_{j}$ is the sample proportion in category $j$. The matrix
$R^{c}=M^{c}\hat{Q}^{c}$ is invariant, that is, $p^{c}R^{c}=p^{c}$,
since $R^{c}_{ij}=\sum_{k}[p^{c}_{j}M^{c}_{ik}M^{c}_{jk}/\break\sum
_{l}M^{c}_{lk}p^{c}_{l}]$ and $\sum_{i}p^{c}_{i}R^{c}_{ij}=\sum
_{k}p^{c}_{j}M^{c}_{ik}=p^{c}_{j}$. The vector of the original
proportions $p^{c}$ is the eigenvector of $R$. In practice, $\hat
{Q}^{c}$ can be calculated by transposing matrix $M^{c}$, multiplying
each column $j$ by $p^{c}_{j}$ and then normalizing its rows so that
the sum of each row equals one. We define $R^{{c}^{*}}=\alpha
R^{c}+(1-\alpha)I$, where $I$ is the identity matrix of the appropriate
size. $R^{{c}^{*}}$ is also invariant and the amount of
misclassification is controlled by the value of $\alpha$.

We conduct a second experiment using the same data and setup described
in Section~\ref{ss} and PRAM to perturb the geographical variable LAD.
For the random perturbation, an $11\times11$ misclassification matrix
$M^{c}$ is defined for the 11 categories of LAD where the diagonal
elements are 0.9 and 0.8 and the off-diagonal elements are equal to a
probability of 0.1 and 0.2 for the $10\%$ and $20\%$ perturbation
respectively. The invariant misclassification matrix is calculated with
$\alpha=0.55$. For each individual, a random uniform number between 0
and 1 is generated and the category of the LAD changed (or not changed)
if it is in the interval defined by the cumulative probability. For the
$10\%$ targeted perturbation, we define the misclassification matrix
for the ``Other'' ethnicities with 0.25 on the diagonal and 0.75 on the
off-diagonals and the invariant parameter $\alpha=0.85$. For the $20\%$
targeted perturbation, we define the misclassification matrix for the
``Other'' ethnicities with 0 on the diagonal and 1 on the off-diagonals,
and the misclassification matrix for the ``White British'' ethnicity with
0.93 on the diagonal and 0.07 on the off-diagonals. For both matrices,
the invariant parameter is $\alpha=1$.

\subsection{Results of disclosure risk assessment}
Since we know the misclassification matrix $M$ and the true population
counts $F_{j}$ in these experiments, we can assess the performance of
expressions (\ref{gh})--(\ref{kl}) as approximations to (\ref{ef}). We
do this by summing all the expressions across sample unique records, as
in the aggregate risk measure $\tau$ in (\ref{vt}) and comparing the
resulting sums. We also compare these measures to the measure in (\ref{ht}) of Gouweleeuw et al. (\citeyear{GouweleeuwEtAl1998}).
In addition, we consider the more practical situation when neither the
$F_{j}$ nor the $\tilde{F}_{j}$ are known to the agency, all that is
observed is the ``misclassified'' sample and the matrix $M$. In this
case, we carry out the risk estimation as described in Section~\ref{zb}
through the use of the Poisson log-linear model on the sample counts
$\tilde{f}_{j}$. The log-linear model was chosen using a forward search
algorithm and the outcome of goodness of fit statistics as developed in
Skinner and Shlomo (\citeyear{SkinnerShlomo2008}). We calculate the naive estimated risk
measure obtained from the log-linear model on the misclassified sample
and the adjusted estimated risk measure, taking into account the
misclassification matrix. The experiments were repeated under different
samples and each perturbation method applied independently and we found
that all of the experiments produced similar results. Table~\ref{parset} presents results of one of the simulation experiments for each
of the perturbation methods: random and targeted data swapping and PRAM.\looseness=1

\begin{table}
\caption{Identification risk estimates for microdata samples generated
from UK 2001 Census subject to perturbative SDL methods---Risk measure $\tau^{*}$ no
misclassification${}={}$358.1}\label{parset}
\begin{tabular*}{\textwidth}{@{\extracolsep{4in minus 4in}}ld{4.1}d{4.1}d{4.1}d{4.1}@{}}
\hline
 & \multicolumn{4}{c@{}}{\textbf{SDL method}}  \\[-6pt]
                    & \multicolumn{4}{c@{}}{\hrulefill}  \\
 & \multicolumn{2}{c}{\textbf{Random}} & \multicolumn{2}{c@{}}{\textbf{Targeted}}
          \\[-6pt]
         & \multicolumn{2}{c}{\hrulefill} & \multicolumn{2}{c@{}}{\hrulefill} \\
\textbf{Identification risk  measures}
& \multicolumn{1}{c}{\textbf{Swap}}
& \multicolumn{1}{c}{\textbf{PRAM}}
& \multicolumn{1}{c}{\textbf{Swap}}
& \multicolumn{1}{c@{}}{\textbf{PRAM}} \\
\hline
\multicolumn{5}{c}{$10\%$ perturbation} \\
\multicolumn{5}{c}{Identification risk measures for perturbed data with known population counts}\\
Risk measure $\tau$ in (\ref{vt}) & 321.6 & 325.8 & 146.3 & 161.6 \\
Approximation in (\ref{gh}) & 321.4 & 325.5 & 146.2 & 161.4 \\
Approximation in (\ref{ij}) & 317.7 & 321.7 & 144.8 & 159.8 \\
Approximation in  (\ref{kl}) & 321.6 & 325.6 & 146.3 & 161.6 \\
Risk measure $\tau^{*}_{\mathit{CC}}$ in (\ref{uv}) & 316.6 & 318.2 & 149.5 &
160.3
\\[3pt]
\multicolumn{5} {c}{Estimated risk measures based on sample data}
\\
Risk measure in  (\ref{ht}) & 2486.7 & 2489.1 & 1749.1 & 1899.3 \\
Naive  risk  measure\\
\quad(Poisson log-linear model\\
\quad on  misclassified  sample) & 343.2 & 347.6 & 297.2 & 285.4 \\
Estimated  risk  measure\\
\quad(Poisson log-linear model\\
\quad adjusted  for  misclassification) & 308.8 & 312.7 & 142.7 & 157.9
\\[6pt]
\multicolumn{5}{c}{20\% perturbation} \\
\multicolumn{5}{c}{Identification risk measures for perturbed data with known population counts}\\
Risk measure $\tau$ in (\ref{vt})  & 298.9 & 299.7 & 82.2 & 133.8 \\
Approximation in  (\ref{gh}) & 298.4 & 299.3 & 82.1 & 133.7 \\
Approximation in (\ref{ij}) & 280.4 & 283.5 & 81.7 & 132.7 \\
Approximation in  (\ref{kl}) & 298.9 & 299.8 & 82.2 & 133.8 \\
Risk measure $\tau^{*}_{\mathit{CC}}$ in (\ref{uv}) & 292.8 & 292.2 & 85.0 & 133.4
\\[3pt]
\multicolumn{5} {c}{Estimated risk measures based on sample data}\\
Risk measure in (\ref{ht}) & 2264.0 & 2311.7 & 1419.8 & 1688.2 \\
Naive  risk  measure\\
\quad(Poisson log-linear model\\
\quad on  misclassified  sample) & 358.6 & 349.5 & 262.5 & 285.2 \\
Estimated  risk  measure\\
\quad(Poisson log-linear model\\
\quad adjusted  for  misclassification) & 286.8 & 283.1 & 90.3 & 133.2 \\
\hline
\end{tabular*}
\end{table}

The estimates presented in Table~\ref{parset} for the risk of
identification are similar for random data swapping and PRAM.
Misclassification reduces the risk in the file from about $\tau
^{*}=360$ to about $\tau^{*}_{\mathit{CC}}=290$ for the $20\%$ perturbation and
$\tau^{*}_{\mathit{CC}}=320$ for the $10\%$ perturbation for those methods. The
measure $\tau^{*}$ is interpreted as the expected number of correct
matches which an intruder would make if matches were attempted with all
sample unique records. The decrease in this measure from 360 to 290 as
a result of misclassification is modest since a large number of records
remain unchanged.
An alternative interpretation of $\tau$ could be obtained by dividing
by the number of sample uniques to give the proportion of sample
uniques which would be expected to be identified correctly. This
proportion ranges in Table~\ref{parset} between 0.116 for the $10\%$
Random Swap, 0.053 for the $10\%$ Targeted Swap, 0.108 for the $20\%$
Random Swap and 0.030 for the $20\%$ Targeted Swap.

The identification risk is reduced considerably with the targeted data
swapping since many more sample uniques are perturbed. The
misclassification is reduced from about $\tau^{*}=360$ to about $\tau
^{*}_{\mathit{CC}}=85$ for data swapping and $\tau^{*}_{\mathit{CC}}=130$ for PRAM for
the $20\%$ perturbation and to about $\tau^{*}_{\mathit{CC}}=150$ for data
swapping and $\tau^{*}_{\mathit{CC}}=160$ for PRAM for the $10\%$ perturbation.
The three approximations to the risk measure in (\ref{vt}) all provide
good results, although the approximation in (\ref{ij}) slightly
underestimates. The measure in (\ref{vt}) relies on knowledge of both
the full misclassification matrix $M$ and the population counts
$F_{j}$. In contrast, the approximations (\ref{gh}), (\ref{ij}) and (\ref{kl}) only require knowledge of the probability of not misclassifying a
record, that is, the probabilities on the diagonals. The alternative
risk measure $\tau^{*}_{\mathit{CC}}$ in (\ref{uv}) also turns out to behave
similarly to (\ref{vt}).
The value of the measure in (\ref{ht}) of Gouweleeuw et al. (\citeyear{GouweleeuwEtAl1998}) is
much higher than the values of the other measures, reflecting the very
conservative assumption that the intruder knows that the target unit is
in the microdata sample.
In practice, the population counts will generally be unknown to the
statistical agency (and the intruder) for survey data. We therefore
consider the method in Section~\ref{zb} based upon the Poisson
log-linear model. The estimated aggregate risk measures appear to
perform well with estimates for the risk measure under
misclassification of about 285 for random data swapping and PRAM under
the $20\%$ perturbation and about 310 for random data swapping and PRAM
under the $10\%$ perturbation. The estimated aggregate risk measures
are about 140 for targeted data swapping and 160 for targeted PRAM for
the $20\%$ perturbation and about 90 for targeted data swapping and 130
for targeted PRAM for the $10\%$ perturbation.

\begin{figure}

\includegraphics{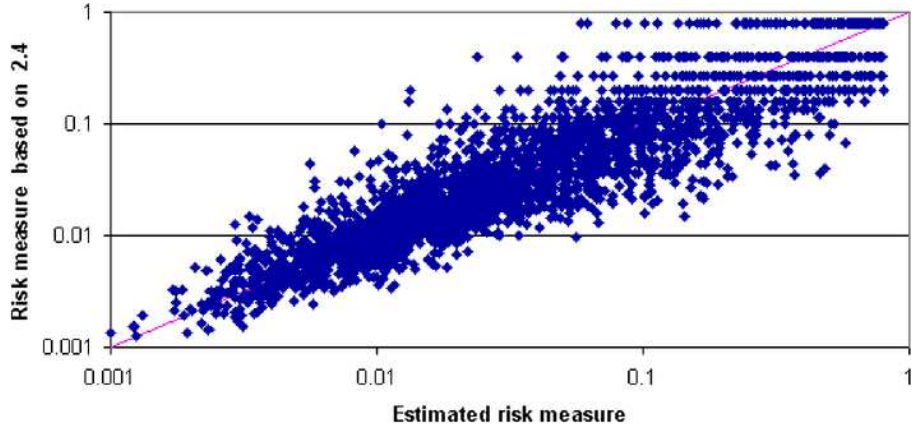}

\caption{Scatterplot of inidividual per-record risk measures in (\protect\ref{ef})
against estimated risk measures based on the Poisson log-linear model
under $20\%$ random data swap.}\label{fig2}
\end{figure}

\begin{table}[b]
\caption{Cross-classification of sample uniques according to per-record risk
measures in (\protect\ref{ef})
and estimates based on Poisson log-linear model under 20\% random data swap}\label{parsee}
\begin{tabular*}{\textwidth}{@{\extracolsep{4in minus 4in}}ld{4.0}d{3.0}d{3.0}d{4.0}@{}}
\hline
& \multicolumn{4}{c@{}}{\textbf{Estimates from Poisson log-linear model}}\\[-6pt]
& \multicolumn{4}{c@{}}{\hrulefill}\\
\textbf{Per-record risk measures from~(\ref{ef})}
& \multicolumn{1}{c}{\textbf{0.00--0.09}}
& \multicolumn{1}{c}{\textbf{0.10--0.49}}
& \multicolumn{1}{c}{\textbf{0.50--1.00}}
& \multicolumn{1}{c@{}}{\textit{\textbf{Total}}} \\
\hline
0.00--0.09 & 1961 & 133 & 4 & 2098 \\
0.10--0.49 & 180 & 325 & 76 & 581 \\
0.50--1.00 & 8 & 69 & 75 & 152 \\
\textit{Total}   & 2149 & 527 & 155 & 2831 \\
\hline
\end{tabular*}
\end{table}

Another important consideration when assessing disclosure risk for
releasing microdata is the individual per-record (record-level)
disclosure risk measures in (\ref{ef}). Individual records with high
disclosure risk might be subjected to further tailored perturbation. In
Figure~\ref{fig2}, we plot the per-record (record-level) risk measures
in (\ref{ef}) for the sample uniques against the estimated adjusted
risk measures (as described in Section~\ref{zb}) based on the Poisson
log-linear model for the experiment based on $20\%$ random data
swapping. In addition, we summarize this bivariate distribution for the
sample uniques in a two-way table in Table~\ref{parsee}. In both of the
analyses we see a good fit between the risk measures in (\ref{ef}) and
their estimated risk measures. The Spearman's rank correlation is 0.91.

\subsection{Results of information loss assessment}\label{sec4.4}
The utility of microdata that has undergone data masking techniques is
measured here in terms of the closeness of the results of an analysis
based upon the perturbed data compared to the same analysis based upon
the original data. The nature of the results and the type of analysis
depend on user requirements. In general, microdata is multi-purpose and
used by many different users. For this assessment we use the following
three information loss measures reflecting distortions of distributions
in two-way tables, as considered by Gomatam and Karr (\citeyear{GomatamKarr2003}) and
Shlomo and Young (\citeyear{ShlomoYoung2006}):

\begin{itemize}
\item Relative absolute average distance per cell: Let $D$ represent a
frequency distribution for a two-way table produced from the microdata
and let $D(r,c)$ be the frequency in the cell in row $r$ and column
$c$. The distance metric is
\[
\mathit{RAAD}(D_{\mathit{orig}},D_{\mathit{pert}})=100\times(D_{\mathit{avg}}-\mathit{AAD})/D_{\mathit{avg}},
\]
where the average cell size is defined as
\[
D_{\mathit{avg}}=\sum_{r,c}D_{\mathit{orig}} (r,c)/RC
\]
with $R$ the number of rows and $C$ the number of columns in the table,
and the $\mathit{AAD}$ metric is defined as
\[
\mathit{AAD}(D_{\mathit{orig}},D_{\mathit{pert}})=\sum_{r,c}|D_{\mathit{pert}}(r,c)-D_{\mathit{orig}}(r,c)|/RC
\]
with $\mathit{pert}$ and $\mathit{orig}$ referring to the perturbed and original tables
respectively. The $\mathit{RAAD}$ provides a measure of the average absolute
perturbation per cell compared to the average cell size of the table.

\item Impact on measures of association:
\[
\mathit{RCV}(D_{\mathit{orig}},D_{\mathit{pert}})
=100\times\bigl(\mathit{CV}(D_{\mathit{pert}})-\mathit{CV}(D_{\mathit{orig}})\bigr)/\mathit{CV}(D_{\mathit{orig}}),
\]
where
\[
\mathit{CV}(D)=\sqrt{\chi^{2}/\min(R-1,C-1)}
\]
is Cramer's measure of association, defined in terms of $\chi^{2}$, the
usual Pearson chi-squared statistic for testing independence in the
two-way table. The $\mathit{RCV}$ provides a measure of attenuation of the association.

\item Impact on an ANOVA analysis: another form of bivariate analysis
consists of comparing proportions in a category of a column (outcome)
variable between categories of a row (explanatory) variable. Let
$P^{c}(r)=D(r,c)/\sum_{c}D(r,c)$ be the proportion in column $c$ for
row $r$ and define the between-row variance of this proportion by
\[
\mathit{BV}(P^{c})= \sum_{r}\bigl(P^{c}(r)-P^{c}\bigr)^2/(R-1),
\]
where $P^{c}=\sum_{r}D(r,c)/\sum_{rc}D(r,c)$. The measure of
information loss is
\[
\mathit{BVR}(P^{c}_{\mathit{orig}},P^{c}_{\mathit{pert}})=100\times
\bigl(\mathit{BV}(P^{c}_{\mathit{pert}})-\mathit{BV}(P^{c}_{\mathit{orig}})\bigr)/\mathit{BV}(P^{c}_{\mathit{orig}}).
\]
The $\mathit{BVR}$ provides a measure of attenuation of between group
differences in an ANOVA analysis.
\end{itemize}

\begin{table}
\caption{Information loss measures for microdata samples generated
from UK 2001 Census subject to three
perturbative SDL methods}\label{pars}
\begin{tabular*}{\textwidth}{@{\extracolsep{4in minus 4in}}l d{3.1}d{3.1}d{3.1}d{3.1}@{}}
\hline
& \multicolumn{4}{c@{}}{\textbf{SDL method}}  \\[-6pt]
& \multicolumn{4}{c@{}}{\hrulefill}  \\
& \multicolumn{2}{c}{\textbf{Random}}
& \multicolumn{2}{c@{}}{\textbf{Targeted}} \\[-6pt]
& \multicolumn{2}{c}{\hrulefill}
& \multicolumn{2}{c@{}}{\hrulefill} \\
\textbf{Information loss measures}
& \multicolumn{1}{c}{\textbf{Swap}}
& \multicolumn{1}{c}{\textbf{PRAM}}
& \multicolumn{1}{c}{\textbf{Swap}}
& \multicolumn{1}{c@{}}{\textbf{PRAM}}
\\
\hline
\multicolumn{5}{@{}c@{}}{$10\%$ perturbation} \\
$\mathit{RAAD}$ on $\mathit{LAD}\times{}$ethnicity & 98.5 & 98.1 & 97.4 & 97.2 \\
$\mathit{RAAD}$ on $\mathit{LAD}\times{}$economic activity & 97.0 & 96.9 & 96.1 & 95.8\\
$\mathit{RCV}$ on $\mathit{LAD}\times{}$ethnicity & -9.9 & -10.4 & -13.3 & -12.9 \\
$\mathit{RCV}$ on $\mathit{LAD}\times{}$economic activity & -10.8 & -9.8 & -11.0 &-10.4 \\
$\mathit{BVR}$ on prop. ``White British'' across $\mathit{LAD}$ & -20.9 & -23.8 &0 & 0 \\
$\mathit{BVR}$ on prop. ``Indian'' across $\mathit{LAD}$ & -12.6 & -13.0 & -18.9 & -17.3
\\[3pt]
\multicolumn{5}{@{}c@{}}{$20\%$ perturbation} \\
$\mathit{RAAD}$ on $\mathit{LAD}\times{}$ethnicity & 97.4 & 97.2 & 96.5 & 96.4 \\
$\mathit{RAAD}$ on $\mathit{LAD}\times{}$economic activity & 95.8 & 95.5 & 95.0 & 94.9\\
$\mathit{RCV}$ on $\mathit{LAD}\times{}$ethnicity & -20.4 & -20.4 & -17.8 & -16.9 \\
$\mathit{RCV}$ on $\mathit{LAD}\times{}$economic activity & -18.1 & -17.0 & -16.2 &-14.4 \\
$\mathit{BVR}$ on proportion ``White British'' across $\mathit{LAD}$ & -37.4 &-39.6 & 0 & 0 \\
$\mathit{BVR}$ on proportion ``Indian'' across $\mathit{LAD}$ & -37.5 & -39.1 &-34.2 & -29.5 \\
\hline
\end{tabular*}
\end{table}

Table~\ref{pars} presents results of the information loss measures on
the misclassified samples used in Table~\ref{parset}. We obtain similar
results for the information loss measures when comparing data swapping
and PRAM with an expected improvement under the smaller perturbation
rate of $10\%$. The targeted perturbation shows slight improvements to
the $\mathit{RAAD}$ compared to the random perturbation under both perturbation
rates. The targeted perturbation is generally worse for the $\mathit{RCV}$ and
$\mathit{BVR}$ compared to the random perturbation under the $10\%$ perturbation
rate, but there are slight improvements under the $20\%$ perturbation
rate. The impact on the $\mathit{BVR}$ for other ethnic groups (not shown) was
mixed with most of the ethnic groups following the same pattern of
attenuation as seen for the ``Indian'' ethnic group. There were a few
exceptions due to small sample sizes. For example, we obtained a
positive value for the $\mathit{BVR}$ of ``Chinese'' ethnicity. Overall, the
considerable reduction in disclosure risk achieved by the $20\%$
targeted data swapping in Table~\ref{parset} does not appear to be
offset by any major reduction in utility compared to the other methods.

\begin{figure}

\includegraphics{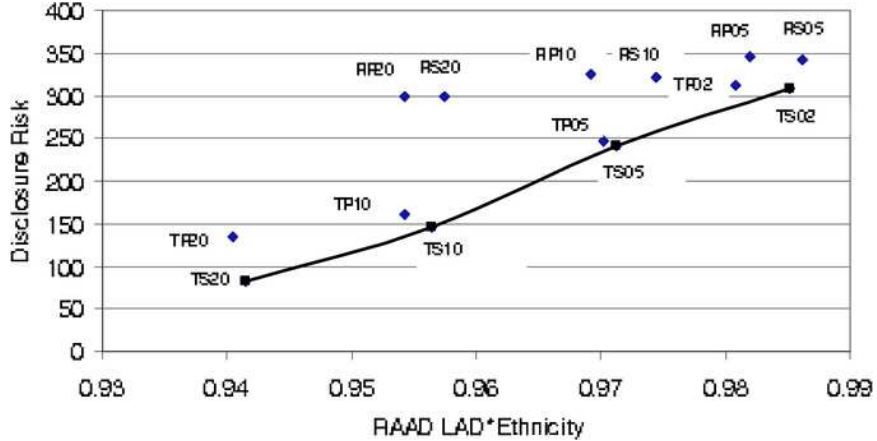}

\caption{Risk-utility map.}\label{fig4}
\end{figure}

In Figure~\ref{fig4} we plot a risk-utility map [Duncan, Keller-McNulty and Stokes (\citeyear{DuncanKellermcnultyStokes2001})].
The points on the map represent different candidate releases, that is,
perturbation methods with different levels of perturbation. In addition
to the levels considered earlier, we also include $2\%$ and $5\%$
targeted and random perturbation. The points are denoted $T$ for
targeted or $R$ for random; 20 for $20\%$, 10 for $10\%$, 5 for $5\%$
or 2 for $2\%$; and S for swapping or P for PRAM. The points are
plotted against the risk measure $\tau$ in (\ref{vt}) on the Y-axis and
the information loss measure $\mathit{RAAD}$ for $\mathit{LAD} \times \mathit{ethnicity}$ on the
X-axis. We see that, at the same level of information loss between the
targeted $10\%$ perturbation and the random $20\%$ perturbation with
respect to the $\mathit{RAAD}$, we obtain lower disclosure risk with the
targeted $10\%$ perturbation. The same applies to the targeted $5\%$
perturbation and the random $10\%$ perturbation, with the targeted $5\%
$ perturbation having less disclosure risk than the random $10\%$
perturbation at the same level of information loss. We draw a line to
connect points on the risk-utility frontier [Gomatam, Karr and Sanil (\citeyear{GomatamKarrSanil2005})] and
note that in all cases, at given levels of information loss, the
targeted data swapping provides the lowest disclosure risk compared to
the other methods, although there is little difference between targeted
swapping and targeted PRAM. Targeting did not appear to lead to much
greater information loss for the other measures in Table~\ref{pars} and the
general conclusion here is that targeting seems useful, enabling less
perturbation to be applied and hence less information loss for a given
level of risk protection. Of course, this finding could vary in other
settings and an agency could use a similar risk-utility approach, based
on its own data, to determine its preferred SDL approach.

\section{Impact of misclassifying multiple key variables}\label{zq}

The previous section only provided estimates of the impact of
misclassifying one key variable. In this section we provide a further
numerical illustration to demonstrate the potential impact of
misclassifying multiple key variables. We consider a simple setup where
the $C$ key variables $X^{1},\ldots,X^{C}$ are independent and binary.
Their values in the external information and the microdata are denoted
$X^{c}$ and $\tilde{X}^{c}$ respectively, $c=1,\ldots,C$. We suppose
that $\operatorname{Pr}(X^{c}=2)=p$, $\operatorname{Pr}(X^{c}=1)=1-p$, $\operatorname{Pr}(\tilde
{X}^{c}=2|X^{c}=1)=\theta_{1}$ and $\operatorname{Pr}(\tilde{X}^{c}=1|X^{c}=2)=\theta
_{2}$ for $c=1,\ldots,C$. The misclassification probabilities $M_{jk}$
in (\ref{ab}) will thus consist of products of $C$ terms, each term
being one of $\theta_{1}$, $1-\theta_{1}$, $\theta_{2}$ or $1-\theta
_{2}$. To force $X$ and $\tilde{X}$ to share the same marginal
distribution, we set $\theta_{2}=(1-p)\theta_{1}/p$ so that $\operatorname{Pr}(\tilde
{X}^{c}=1)=p$ and, to simplify, write $\theta_{1}=\theta$.

In our experiment we generated values of $X$ for a population of size
$N$, drew a sample of size $n$ by simple random sampling and then
generated the values $\tilde{X}$. Various choices of $(N,n,C,p,\theta)$
were considered. We also generated $\tilde{X}$ for all population units
so that $\tilde{F}_{j}$ could be computed.

\begin{figure}

\includegraphics{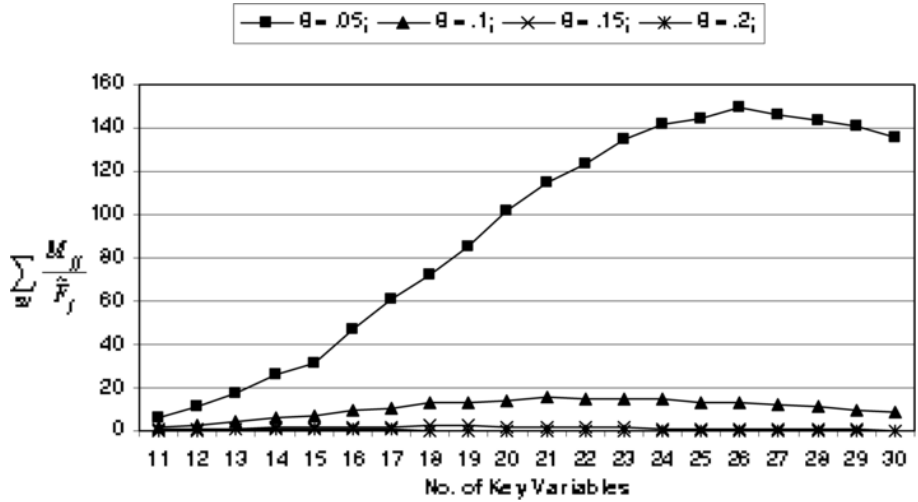}

\caption{Risk measure for different numbers of key variables and rates
of misclassification.}\label{fig1}
\end{figure}

We report values of risk measure (\ref{gh}) summed over sample uniques
$\sum_{\mathit{SU}}M_{jj}/\tilde{F}_{jj}$ in Figure~\ref{fig1} for
$N=100{,}000$, $n=2000$, $p=0.2$ and for various choices of $C$ and
$\theta$. Note that the number of sample uniques increases as we add in
more binary key variables. For $C=11$ we have about 240 sample uniques
and for $C=30$ we have about 1960 sample uniques.
In the absence of misclassification, we find that the risk increases
monotonically and rapidly with $C$. This is because the number of
population uniques is increasing with $C$ and the fact that any
observed match with a population unique must be a true match. On the
other hand, in the presence of misclassification, we find that the risk
does not increase monotonically, rather it reaches a maximum and then
declines. As expected, the more misclassification, the lower the
disclosure risk.

We do not present information loss measures for the simulation since
their values follow theoretically. For any analysis involving a given
set of variables, say, the estimation of a table cross-classifying two
particular key variables, the addition of further key variables will
have no systematic impact on any of the information loss measures,
since each of the variables of interest will be perturbed in the same
way, irrespective of the inclusion of other key variables. The only
variation we might expect to observe would be as a result of simulation
variation. Any information loss function in Figure~\ref{fig1} should therefore
be flat.

\section{Discussion}\label{zs}

In this paper we have shown how existing methods for assessing
identification risk in survey microdata may be extended in a relatively
simple way to capture the impact of SDL methods based on
misclassification. We presented a general expression for the risk under
misclassification in (\ref{ef}) and showed that the simple formula in
(\ref{gh}) provided a good approximation to this expression in two
experiments based upon UK census data. The advantage of the formula in
(\ref{gh}) is that it enables the extension of existing risk assessment
methods for unpeturbed data based on Poisson log-linear models, as
discussed in Skinner and Shlomo (\citeyear{SkinnerShlomo2008}), to handle perturbative SDL
methods. We demonstrated this extended approach also with the census
data and provided a disclosure risk-data utility analysis. We showed
how a targeted SDL method could dominate corresponding random methods.

One challenge faced by agencies when assessing identification risk is
the need to make assumptions about the information available to the
intruder, specifically the nature and number of key variables. We
conducted a numerical experiment to assess the sensitivity of the
identification risk to the misclassification of different numbers of
key variables. In the absence of misclassification, the risk can
increase rapidly with the number of key variables. We observed that
misclassification can, however, dominate this effect with the risk
eventually declining as the number of key variables increases. This is
potentially an encouraging finding for agencies, since the sensitivity
of the identification risk to departures from assumptions about the
choice of key variables may be reduced in some settings when the kinds
of SDL methods considered here are used and, in cases such as in
Figure~\ref{fig1}, there may even be a natural upper bound for the risk
across plausible choices.

Another issue faced by agencies is whether to release values of the
parameters of the SDL method employed, for example, the swapping rate.
The information loss measures used in Section~\ref{sec4.4} assume that users of
the microdata simply ignore the perturbation in their analyses of the
data. The agency's aim is to find an SDL method for which both the
information loss and the disclosure risk are considered satisfactorily
small. If this is not feasible, then it may be necessary for the agency
to resort to an SDL method which leads to nonnegligible distortion of
analyses. In this case it may be desirable for data analysts to be
provided with values of the parameters of the SDL method to enable them
to undertake valid inference, as discussed, for example, in
Gouweleeuw et al. (\citeyear{GouweleeuwEtAl1998}) for PRAM (note that our use of invariant PRAM was
designed to avoid this need). The disclosure risk implications of
releasing such SDL parameters will not be pursued further here.

The findings of this paper are not only relevant to understanding the
impact of SDL methods, but also to the assessment of risk, before the
application of SDL methods, in a way which more realistically takes
account of the errors of classification which arise in survey data from
measurement, coding and processing as well as from imputation for
missing data, providing the agency has estimates for the diagonal
elements of the misclassification matrix.

\begin{appendix}
\section*{\texorpdfstring{Appendix: Derivation of (\protect\lowercase{\ref{gh}}) under probabilistic\break record
linkage}{Appendix: Derivation of (2.5) under probabilistic\break record
linkage}}\label{appendix}

Suppose, as before, that a microdata record $i$ is linked to a target
unit $B$ by comparing the values of $\tilde{X}_{i}$ and $X_{B}$.
Following the approach of Fellegi and Sunter (\citeyear{FellegiSunter1969}), let $\gamma(\tilde
{X}_{i},X_{B})=j$ if
$\tilde{X}_{i}=X_{B}=j$, $j=1,\ldots,K$, and $\gamma(\tilde
{X}_{i},X_{B})=K+1$ if
$\tilde{X}_{i}\neq X_{B}$ and suppose that exact matching is used, so
that a link is made if $\gamma(\tilde{X}_{i},X_{B})\leq K$. Suppose the
intruder draws the pair $(i,B)$ at random (with equal probability) from
the set of
pairs $s\times s^{\ast}$, where $s^{\ast}$ is the subset of units in
$U$ appearing in the external database from which the intruder
selects $B$.
Partition $s\times s^{\ast}$ as $M=\{(i,B)|i=B\}$ and $U=\{(i,B)|i\neq B\}$
and let $m(j)=\operatorname{Pr}[\gamma(\tilde{X}_{i},X_{B})=j|(i,B)\in M],$ $u(j)=\operatorname{Pr}
[\gamma(\tilde
{X}_{i},X_{B})=j|(i,B)\in U]$ and $p=\operatorname{Pr}[(i,B)\in M],$ where $\operatorname{Pr}(\cdot)$ is
defined with respect to the selection of $(i,B)$, the selection of the sample
$s$ and the misclassification mechanism. Then the identification risk
for a
linked pair $(i,B)$ for which $\tilde{X}_{i}=X_{B}=j$ is given by
\[
\phi_{j}=\operatorname{Pr}[(i,B)\in M|(\tilde{X}_{i},X_{B})=j]=\frac{m(j)p}{m(j)p+u(j)(1-p)}.
\]

A large sample size approximation gives $m(j)\approx M_{jj}f_{j}/n^{\ast
}$, $u(j)\approx(\pi\tilde{F}_{j}f_{j}-\pi M_{jj}f_{j})/(nn^{\ast}-\pi
n^{\ast})$, $p=\pi/n$, where $f_{j}$ is the number of units $b$ in
$s^{\ast}$ for
which $X_{b}=j$ and $n^{\ast}$ is the size of $s^{\ast}$. It follows
that $\phi_{j}\approx M_{jj}/\tilde{F}_{j}$
irrespective of the manner in which $s^{\ast}$ is selected from $U$.
Skinner (\citeyear{Skinner2008}) provides further discussion of identification risk under probabilistic
record linkage.
\end{appendix}

\printaddresses

\end{document}